\documentclass[preprint2]{aastex}
\shorttitle{Detection of a Companion Lens Galaxy}
\shortauthors{MacLeod, Kochanek, \& Agol}
\begin{document}
\title{Detection of a Companion Lens Galaxy using the Mid-infrared Flux Ratios 
of the Gravitationally Lensed Quasar H1413+117}

\author{Chelsea L. MacLeod\altaffilmark{*}, Christopher S. Kochanek\altaffilmark{**},
  and Eric Agol\altaffilmark{*}}
\altaffiltext{*}{Astronomy Department, University of Washington, Seattle, WA
  98195}
\altaffiltext{**}{Department of Astronomy and the Center for Cosmology and
  Astroparticle Physics, The Ohio State University, Columbus, OH
  43210}

\begin{abstract}
We present the first resolved mid-IR (11$\mu$m) observations of the four-image
quasar lens H1413+117 using the Michelle camera on
Gemini North. All previous observations (optical, near-IR, and radio)
of this lens show a ``flux anomaly,'' where the image flux ratios
cannot be explained by a simple, central lens galaxy.  We attempt to
reproduce the mid-IR flux ratios, which are insensitive to extinction and
microlensing, by modeling the main lens as a singular isothermal
ellipsoid. This model fails to reproduce the flux ratios. However, we can
explain the flux ratios simply by adding to the model a nearby galaxy
detected in the H-band by HST/NICMOS-NIC2.
This perturbing galaxy lies $4\farcs0$ from the main lens and it 
has a critical radius of $0\farcs63\pm0\farcs02$ which is 
similar to that of the main lens, as expected from their similar H-band fluxes. More 
remarkably, this galaxy is not required to obtain a good
fit to the system astrometry, so this represents the first clear detection
of an object through its effect on the image
fluxes of a gravitational lens. This is a parallel to the detections of visible satellites
from astrometric anomalies, and provides a proof of the concept of
searching for substructure in galaxies using anomalous flux ratios.  
\end{abstract}


\section{Introduction}
Ever since cosmological simulations of cold dark matter (CDM) indicated there
should be many more satellite galaxies around the Milky
Way than detected, it has been an important goal to try to detect
these ``missing satellites'' \citep{kly99,moo99}. In the case of the
Milky Way, the Sloan Digital Sky Survey has steadily found additional, faint satellites,
but with nowhere near the expected abundances of subhaloes
\citep[e.g.][]{will05,bel07}. The simplest solution in the context of CDM
models is to suppress star formation in low-mass satellites, probably
through heating and baryonic mass loss as the universe re-ionizes to
leave a population of dark satellites
\citep[e.g.][]{kly99,bul00}. Locally there is some hope of finding these dark
satellites through gamma rays emitted by dark matter annihilation, but
the likelihood of detection depends heavily on the CDM properties
\citep[e.g.][]{str08}. 

Currently the only other technique able to search for dark substructures is
gravitational lensing, where satellites affect both the image positions and fluxes. In some lenses (e.g. MG0414+0534, \citealt{ros00};
MG2016+112, \citealt{mey06,mor09}; HE0435--1223, \citealt{koc06}) satellite galaxies of the primary lens can
be detected through their effects on image positions.  But, as
\citet{mao98} and \citet{met02} emphasized, image fluxes are sensitive to
perturbations from very low-mass satellites, and we observe many lenses with ``flux
ratio anomalies'' where the relative image brightnesses cannot be
explained by simple, central lens galaxies \citep[see][]{eva03,koc04,con05,yoo06,yoo06b}. Image
fluxes can be altered by 
two physical effects: granularity in the gravitational field created
by stars \citep[``microlensing,'' see][]{wam06} or satellites (``millilensing''), and
propagation effects in the ISM of the lens galaxy
(scattering/extinction).  At radio frequencies, where absorption is
negligible and the sources are too large to be significantly
affected by microlensing, a simple explanation for the anomalous
flux ratios in four-image radio lenses is perturbations due to the
expected CDM substructure. Nothing has altered this
basic conclusion (see, e.g.\
\citealt{dal02,met02,koo03,eva03,con05,yoo06}; Dobler \& Keeton 2006),
although there is debate about whether some of the perturbers may be
along the line of sight \citep{che03,wam05,met05} and about the fraction of substructures that
survive at the typical impact parameter of lensed images
\citep{zen05,mad08,xu09}. To date, however, no visible satellite or companion galaxy has been convincingly detected using flux ratios
rather than astrometry; therefore, we lack a ``proof of principle''
demonstration of the flux ratio approach. 

The need to fully disentangle the physical effects is illustrated 
by the fact that one optical anomaly included by 
\citet{dal02}, that of PG1115+080, was shown by the mid-IR observations of \citet{chi05}
to be due to microlensing rather than substructure. Like radio
emission, the mid-infrared is also insensitive to
extinction and microlensing, so it is an ideal bandpass in
which to search for substructure in the lens.  Quasar spectra rise longward
of 1$\mu$m in $\nu F_{\nu}$ due to a dust ``torus'' surrounding the
AGN \citep{ant93}, and the mid-IR emission of the torus is too large ($\sim$0.1~mas) to be
affected by stellar microlensing.  In addition, the mid-IR source should have a
variability time scale long compared to the lens time delays, leading
to little variability in the mid-IR flux ratios \citep[e.g.][]{poi07}.

In this paper we discuss the lens system H1413+117 \citep{mag88}, for which we have
obtained mid-IR (11$\mu$m) data with the Michelle camera \citep{roc04} on Gemini North. H1413+117 is a four-image quasar lens
with a quasar redshift $z_s=2.55$. The lens galaxy is faint and difficult to detect against
the glare from the bright quasar. There are position estimates from 
\citet{kne98} and \citet{cha07} based on the same HST/NICMOS-NIC2 H-band
image, but no lens redshift. There are many published models of
H1413+117 (\citealt{kay90,yun97,kne98a}; Chae \& Turnshek 1999) using a broad range of mass distributions. The models have never had difficulty matching the image positions, but all have failed to reproduce the flux ratios.

A difficulty for all these previous studies is that the image fluxes are likely affected by both extinction
\citep{tur97,fal99} and (chromatic?) microlensing
(see \citealt{ost97,ang90,cha01}; Hutsemekers 1993). Our 11$\mu$m data essentially
eliminate both of these complications. For a \citet{car89} $R_V=3.1$ extinction
curve, the absorption at $11\mu$m compared to that at H-band is only
$R_{11}/R_H \simeq 0.05$, approaching $0.07$ only as
the lens redshift approaches that of the source. \citet{tur97}
estimated a maximum extinction for H1413+117
relative to image C of approximately $0.1 < A_H < 0.3$, depending on the
assumed dust model and redshift, corresponding to negligible
corrections ($A_{11} < 
2$\%) at 11$\mu$m given our measurement uncertainties. 
In the quasar rest frame, $11\mu$m corresponds to $3\mu$m, a
wavelength where the emission should be dominated by the large dust
torus \citep{san89}.  For microlensing by stars
in the lens galaxy (at $z_{lens}=1$), the Einstein radius of the stars
is $10^{16} \langle M/0.25M_\odot\rangle^{1/2}$~cm, so
the torus radius
is roughly 100 times larger than the Einstein radius of the
typical star and microlensing effects will be weak
\citep[e.g.][]{ref91}. 

The particular model we advance here was considered by
\citet{cha99}, where it had difficulty reproducing the optical flux
ratios. We improve on their results by presenting models based on mid-IR flux ratios
with their insensitivity to extinction and microlensing.  In
Section~\ref{sec:obs} we present our mid-IR observations of H1413+117. In
Section~\ref{sec:mod} we present a simple model of the lens system
that reproduces the observed flux ratios based on perturbations from a
faint companion galaxy.  This galaxy is not required to obtain a good
fit to the system astrometry, so this represents the first clear detection
of an object (albeit a visible object) through its effect on image
fluxes. This is a parallel to the detections of visible satellites
from astrometric anomalies. In Section~\ref{sec:summ} we discuss
our results.

\section{Flux measurements and errors}
\label{sec:obs}
\begin{figure*}[t!]
   \centering
\includegraphics[width=6.0in,viewport=0 0 510 179,clip]{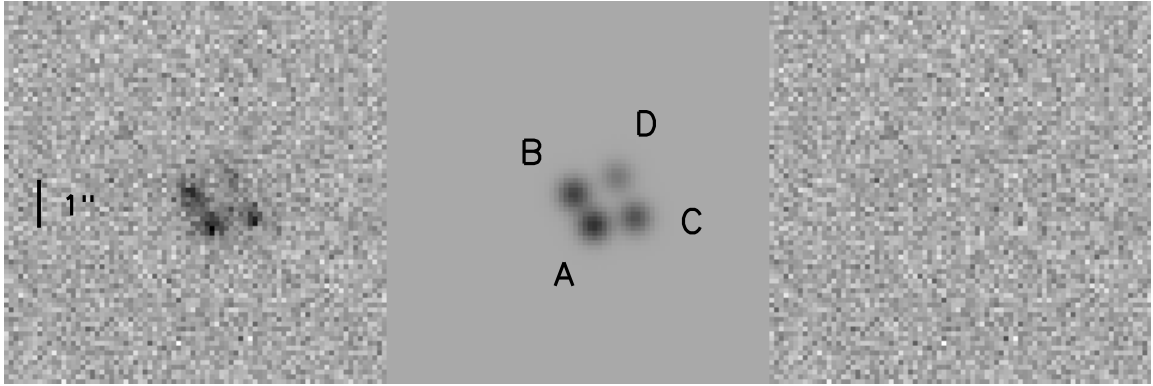}
\caption{\footnotesize The left, middle, and right panels show the 11 micron
  data, the best-fit model based on 2-D
  Gaussians, and the model residuals, respectively.}
\label{fig:images}
\end{figure*}

We observed H1413+117 with the
Michelle camera on Gemini North at 11.2 microns (F112B21 filter) on 5
Aug 2005. The observing time was 1/2 hour, of
which 493.6 seconds were spent on source. The data were processed with the standard Gemini
pipeline {\it mireduce}. Our analysis starts with the
coadded chop and nod subtracted image. These initial images have 16 vertical stripes due to the 16 readout channels; we
corrected for this by subtracting the median of each stripe, masking
the region containing the quasar images. This procedure also removes
any residual sky flux.  Figure~\ref{fig:images} shows the resulting image. 

\begin{figure}[h]
  \centering
    \epsscale{1}
    \plotone{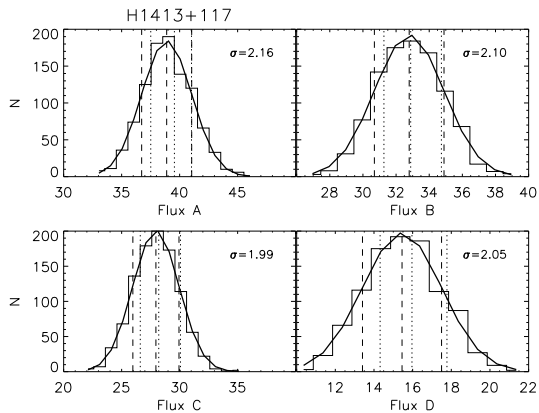}
\caption{\footnotesize Probability distribution from the MCMC method for the
  H1413+117 image fluxes (which are in arbitrary units).
 The mean and $1\sigma$ error bars determined from the
  best-fit Gaussian curve are shown by the vertical dashed lines.
  These are compared to the best-fit flux value and 68\% confidence limits (or parameter values at which
  $\Delta\chi^2 = 1$ from the minimum $\chi^2$) that result from the $\chi^2$ minimization
  technique (see text), which are shown by vertical dotted lines.
  The Gaussian width ($\sigma$) is listed in the top right corner and is used in our final results.}
\label{fig:clovererrors}
\end{figure}

We fit the four lensed images using a 2-D Gaussian to
model the PSF, fixing the relative image positions to those measured
from HST images \citep{tur97}. Since the sky is so much brighter than the quasar images, we use the
standard deviation of the background pixels as an estimate of the errors of the flux of
each pixel.  The model has nine free
parameters:  the fluxes of the four images (4), the position of image A (2), the
standard deviation of the Gaussian PSF along the two principal directions (2),
and the rotation angle of the Gaussian PSF.  We found the best-fit model using
non-linear Levenberg-Marquardt optimization.  We used a Markov chain
Monte Carlo (MCMC) with 10$^4$ steps to compute the parameter
uncertainties, and we have run multiple
chains to check the convergence of the best-fit parameters and their
uncertainties.  We also estimated the errors by simply computing the
goodness of fit as a function of each image flux, marginalizing over
all other variables.  Figure~\ref{fig:images} shows the best-fit model
and its residuals,
and Figure~\ref{fig:clovererrors} shows the MCMC and $\Delta\chi^2$
estimate for the fluxes (in arbitrary units) and their uncertainties.
We calibrated the flux measurements using observations of the standard star 
20 Boo and obtained a total flux for H1413+117 consistent with the value
from \citet{aus98} of $\sim30$~mJy. The image flux ratios (relative to image A) are
listed in Table~\ref{tab:errors}. 

\begin{deluxetable}{c c c c}
\tablewidth{0pt}
\tablecaption{H1413+117 Image Flux Ratios. \label{tab:errors}}
\tablehead{
Image & mid-IR & Kneib et al. &  Chae \& Turnshek }
\startdata
A & $\equiv 1.00$    & $\equiv 1.00$   & $\equiv 1.00$  \\
B & 0.84  $\pm$ 0.07 & 0.92 $\pm$ 0.04 & 0.96 $\pm$ 0.10\\
C & 0.72  $\pm$ 0.07 & 0.71 $\pm$ 0.03 & 0.59 $\pm$ 0.10\\
D & 0.40  $\pm$ 0.06 & 0.57 $\pm$ 0.02 & 0.40 $\pm$ 0.06\\
\tableline
\enddata
\footnotesize
\tablecomments{The left column shows our best-fit mid-IR flux ratios and
  1-sigma MCMC errors. The other columns show corresponding ratios from
  HST/NICMOS-NIC2 H-band \citep{kne98} and HST optical/UV measurements
  after corrections for extinction and microlensing \citep{cha99}.}
\end{deluxetable}

Table~\ref{tab:errors} also compares our flux ratios to the near-IR
values derived from the HST/NICMOS-NIC2 H-band image by \citet[whose values agree with \citealt{cha07}]{kne98}
 as well as the values adopted
by \citet{cha99} based on optical/UV HST photometry with
corrections for extinction (assuming SMC-like dust at redshift 2.55)
and for microlensing in image D (based on the line-to-continuum flux ratios
of \citealt{hut93}). Our flux ratios are significantly inconsistent with
   \citet{kne98}, with a $\chi^2$ of $8.2$ for the two measurements
   agreeing, and mildly inconsistent with \citet{cha99},
   with a $\chi^2$ of $2.1$.  The differences
   are dominated by images D and B for \citet{kne98}
   and images C and B for \citet{cha99}. These
   characterizations of the differences include the
   uncertainties in both measurements, but given our
   uncertainties, the central values of \citet{kne98}
   and \citet{cha99} are ruled out
   with $\chi^2=9.4$ and $6.4$, respectively.

\section{Models}
\label{sec:mod}
With the intrinsic flux ratios in hand, where by ``intrinsic'' we mean
independent of microlensing and ISM absorption, we can now
explore the origin of any flux ratio
anomalies. We model the system with LENSMODEL\footnote{http://redfive.rutgers.edu/$\sim$keeton/gravlens/}
\citep{kee01} using the HST WFPC/WFPC2 image positions
(\citealt{tur97}, see
Table~\ref{tab:pos}). We considered three estimates of the lens
position, all derived from the same HST/NICMOS-NIC2 F160W (H-band)
image: that from \citet{kne98}, that from \citet{cha07}, and our own estimate. The three estimates (see Table~\ref{tab:pos}) differ,
although our estimate is more consistent with \citet{cha07}.  For our
standard model we use a relatively loose constraint on the lens position of
$(\Delta\alpha,\Delta\delta)=(-0\farcs14\pm0\farcs02,0\farcs56\pm0\farcs02)$,
which encompasses these results and the likely systematic problems in
determining the position of the faint lens.  We also include weak priors on the
ellipticity of the main lens ($1-b/a=0.0\pm 0.5$) and the external shear
($\gamma=0.05\pm 0.05$). The main lens galaxy (G1) was initially modeled as a singular isothermal
ellipsoid combined with an external shear.  This is the simplest plausible
model for a lens.  There is considerable evidence
in favor of the isothermal profile, and properties of 4-image
systems other than
time delays are very insensitive to the radial mass
distribution \citep[see][]{mey06}.

\begin{deluxetable}{c c c}
\tablewidth{0pt}
\tablecaption{H1413+117 Image and Lens Positions. \label{tab:pos}}
\tablehead{
 Component & $\Delta\alpha$ ($\arcsec$) & $\Delta\delta$
 ($\arcsec$)}  
\startdata
 A     & $\phantom{+}0.000\phantom{0}\pm0.003\phantom{0}$&$0.000\phantom{0}\pm 0.003\phantom{0}$\\
 B     & $-0.744\phantom{0}\pm0.003\phantom{0}$&$0.168\phantom{0}\pm 0.003\phantom{0}$\\
 C     & $\phantom{+}0.492\phantom{0}\pm0.003\phantom{0}$&$0.713\phantom{0}\pm 0.003\phantom{0}$\\
 D     & $-0.354\phantom{0}\pm0.003\phantom{0}$&$1.040\phantom{0}\pm 0.003\phantom{0}$\\
 G1a& $-0.112\phantom{0}\pm0.02\phantom{00}$ &$0.503\phantom{0}\pm 0.02\phantom{00}$\\
 G1b& $-0.1365\pm0.0024$ &$0.5887\pm 0.0035$\\
 G1c& $-0.142\phantom{0}\pm0.02\phantom{00}$ &$0.561\phantom{0}\pm 0.02\phantom{00}$\\ 
\tableline
\enddata
\footnotesize\tablecomments{Negative RA values are Eastward of Image A.  Image positions are
  taken from \citet{tur97}.  All three lens position estimates,  G1a
  \citep{kne98}, G1b \citep{cha07}, and G1c (our analysis) are
  based on the same HST/NICMOS-NIC2 F160W image.}
\end{deluxetable}

This simple model has no difficulty
fitting the image positions,
but cannot reproduce the mid-IR flux
ratios. We find similar results when using
the lens positions of \citet{kne98} or \citet{cha07}.  If we use
    the very small position errors for G1 from the latter study
    ($0\farcs004$), the fit is dominated by the lens position
    ($\chi^2_{G1}\simeq 40$), but the fluxes are still poorly
    fit.  We feel that our error estimates for the position of the
    lens galaxy (found by bootstrap
    resampling the dithered sub-images and refitting the model)
    are likely to be more realistic. For our standard model, we find
$\chi^2=21.4$ for $DOF=6$ degrees of
freedom.  The models have 15 constraints (the positions of 4 lensed
images and the lens galaxy, the 3 flux ratios, plus the priors, albeit
weak, on the ellipticity $e$ and shear $\gamma$) and 9 free
parameters (the position, mass scale $b_{G1}$, ellipticity $e$, position
angle $\theta_e$ of G1, the amplitude $\gamma$ and
position angle $\theta_{\gamma}$ of the external shear, plus the
source position).  Table~\ref{tab:2galmodelsH1413} presents the
best-fit model parameters and a breakdown of the contributions to the
goodness of fit. We examined whether the problem was specific to
a particular image by sequentially broadening the flux error bars for
each image.  We find improvements in the $\chi^2$ of 11.5, 0.2, 1.3, and 8.8
when we relax the flux constraints on images A, B, C, and
D, respectively, suggesting that the problem lies in the fluxes for
the saddle point images A and D.

\begin{figure*}[t!]
\centering
\epsscale{1.8}
    \plottwo{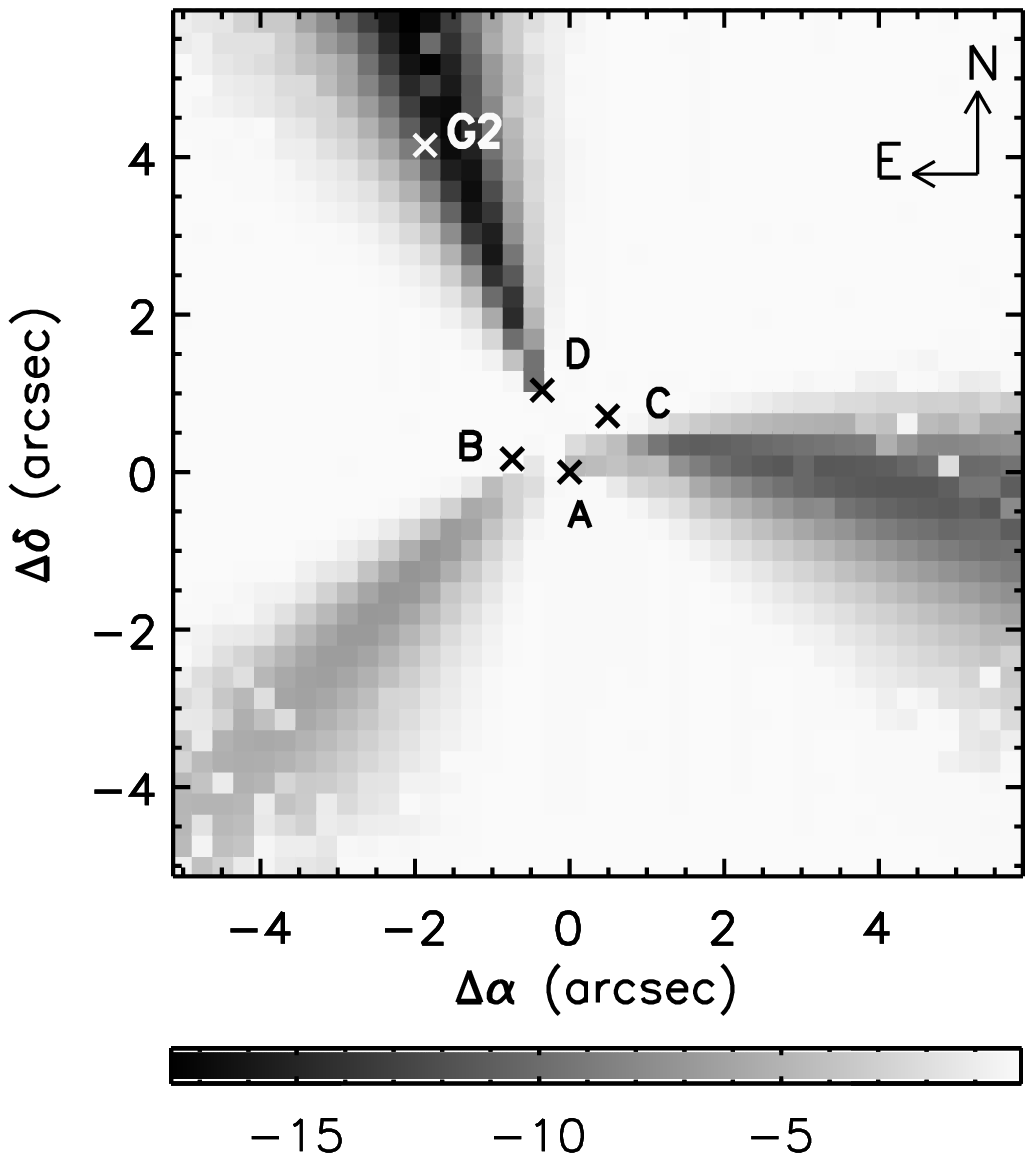}{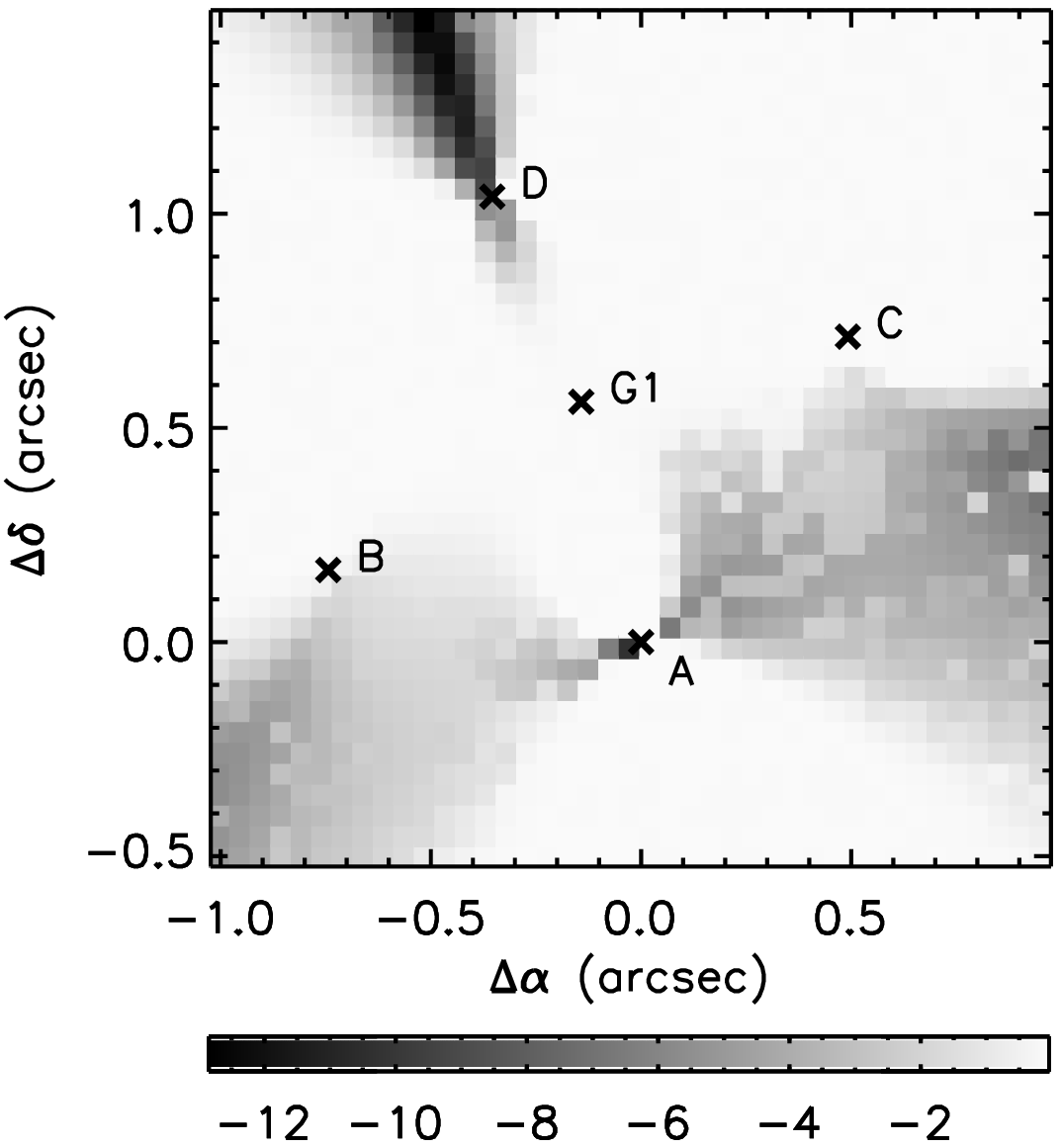}
\caption{\footnotesize The left panel shows the $\Delta\chi^2$ resulting from
  varying the position of a second lens galaxy. The right panel
  is a similar grid focused on the main lens region. Only limited effort was expended to force the 
convergence of recalcitrant pixels.}
\label{fig:chimap}
\end{figure*}

If we now consider a sequence of models adding a singular isothermal
sphere (SIS) as a second lens on a grid of positions, we find
the largest improvement in $\chi^2$ near the position of a galaxy (G2) lying
at $(-1\farcs87,+4\farcs15)$ from image A (see
Figure~\ref{fig:chimap}). This galaxy is object \#14 in \citet{kne98a}
and has an H-band flux of H$=20.0\pm0.2$ mag that is similar to
their estimate for the flux of G1 ($20.6\pm0.5$ mag). This was also
noted in \citet{cha99}, where in their models the flux of image D
was still poorly fit.  If we
simply add an SIS at the observed position of G2, we significantly improve
the fit to $\chi^2=4.9$ for $DOF=5$. We also find better fits to the flux ratios
when using the lens positions of \citet{kne98} or
\citet{cha07}.  G2 is close enough to the lens
(at $r_{G2}=4\farcs0$)
that it contributes more to the model than just an external shear and
convergence. If we attribute the shear in our first model
to G2, then we know that G2
induces a shear of order
$\gamma_{G2} = b_{G2}/(2r_{G2}) \simeq 0.14$ and that it will produce higher
order perturbations beyond the second order 
shear and convergence of order $\gamma_{G2}b_{G1}/r_{G2} \simeq 0.02$.  If we compare
this to the eigenvalues of the magnifications,
$(\lambda_-,\lambda_+)=(-0.15,1.1), (0.20,0.86), (0.20,0.92)$, and
$(-0.24,1.12)$ for images A through D
respectively in our first model, we see that these higher order terms
can significantly alter the fluxes of the images.  This then leads
to the $\Delta\chi^2=16.5$ improvement in the fit.  We also note that
the orientation $\theta_{\gamma}$ of the external shear in our first
model is nearly aligned with the direction of G2. The best-fit parameters of the
two-galaxy model are also reported
in Table~\ref{tab:2galmodelsH1413}.
Note that the critical
radii of G1 and G2, $b_{G1}=0\farcs66$ and $b_{G2}=0\farcs63$, are very similar, as we would expect
from their similar fluxes.  If we vary the mass of G2, we find it is
well-constrained with $b_{G2}=0\farcs63\pm0\farcs02$, despite the freedom to simultaneously adjust the
   external shear.

We investigated whether we had any sensitivity to the mass
  distributions of G1 and G2.  First, we replaced G2 by a point
  mass, and found no significant change in the goodness of fit
  ($\Delta\chi^2=-0.9$).  This is not surprising, as the changes can only come
  from the difference in mass within the annulus about G2 that covers the
  lensed images.  Second, we replaced G1 by an ellipsoidal
  pseudo-Jaffe model ($\rho \propto r^{-2}(r^2+a^2)^{-1}$)
  with the halo truncated at half the Einstein radius of
  the SIE model ($a=0\farcs33$).  This has a series of more
  significant effects.  In order for G1 to produce the same
  quadrupole moment at the Einstein ring, it must have higher
  ellipticity as it becomes more centrally concentrated.
  Moreover, the required shear and ellipticity will scale with the surface
  density at the ring, roughly as $1-\kappa$ \citep{mey06}.  This means
  that a centrally concentrated model for G1 will require
  a flatter model for G1 and a more massive G2, leading to
  changes in the balance between the second order terms
  (shear) and the higher order terms that could be more
  easily detected than when changing the radial structure of G2. While we generally observe
    these effects when we use the truncated model with $a=0\farcs33$,
    the increase to $\chi^2=5.7$ ($\Delta\chi^2=0.8$) indicates that
    we also lack constraints that can distinguish between extended
    and truncated halos for G1.

\begin{figure}[t!]
\epsscale{1}
    \plotone{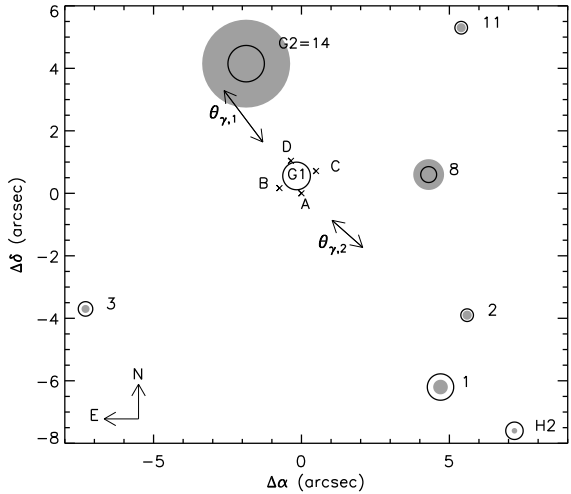}
\caption{\footnotesize Locations of the neighbors of H1413+117 detected in the H-band
  \citep{kne98a}. Galaxies are plotted as open circles with radii
  equal to to their critical radii estimated from H-band magnitudes, and
  as filled circles with sizes scaled to their nonlinear effects listed in the
  last column of Table~\ref{tab:others}. The double-headed arrows indicate the direction of the
  external shear in the single galaxy model ($\theta_{\gamma,1}$) and
  two-galaxy model ($\theta_{\gamma,2}$).}
\label{fig:others}
\end{figure}

Interestingly, the orientation of the external shear in the two-galaxy model,
$\theta_{\gamma}=50\fdg1$, is nearly in the direction of another galaxy in the H-band image \citep{kne98a}, namely object
\#1. This suggests a more general analysis of these neighbors. Figure~\ref{fig:others} shows
positions of all the nearby objects with significant H-band detections
(objects H2, 1, 2, 3, 8, 11, and 14 in \citealt{kne98}). If we assume these
are all in the same group as G1 and scale the Einstein radii of these objects by
their H-band fluxes assuming a standard Faber-Jackson relation ($L
\propto \sigma^4$), we can estimate the deflection scale of each
galaxy near the lens as $b_n=b_{G2}(L_n/L_{G2})^{1/2}$, using the fact that
$b\propto\sigma^2$ for an SIS. Table~\ref{tab:others} lists our
estimates for the critical radii, $b_n$, shear, $\gamma_n=b_n/2r_n$, and the amplitude of any
higher order perturbations, $b_n b_{G1}/r_n^2$, for each galaxy, where we
use our best-fit critical radius of $0\farcs6$ for $b_{G2}$. The galaxies in Figure~\ref{fig:others} are shown by 
open circles representing their critical radii and filled circles
scaled to their higher order perturbations, and the
orientations of the external shear in our single and two-galaxy models are
indicated with arrows.

\begin{deluxetable}{rrrrr}
 \tablewidth{0pt}
\tablecaption{Modeling Results. \label{tab:2galmodelsH1413}}
 \tablehead{
Model:& G1 only & G1+G2 & G1+G2+\#1 & G1+G2+\#1+\#8}
\startdata
$\chi^2/DOF$        & $21.4/6$               & $4.9/5$                     & $3.4/4$                     &$2.9/3$ \\
$\chi^2_{image}$    & 0.3                    & 0.1                         & 0.1                         &0.1\\
$\chi^2_{flux}$     & 15.3                   & 2.4                         & 1.2                         &0.7\\
$\chi^2_{G1}$       & 2.0                    & 1.6                         & 1.8                         &1.8\\
$\chi^2_{prior}$    & 3.8                    & 0.8                         & 0.3                         &0.3\\
$b_{G1}$            & $ 0\farcs69$           & $0\farcs66 $                & $0\farcs63 $                &$0\farcs62$\\
$\Delta\alpha_{G1}$ & $ -0\farcs167$         & $ -0\farcs166$              & $ -0\farcs168$              & $ -0\farcs168$\\
$\Delta\delta_{G1}$ & $\phantom{+}0\farcs549$& $\phantom{+}0\farcs556$     & $\phantom{+}0\farcs555$     & $\phantom{+}0\farcs554$\\
$e$                 & 0.22                   & 0.26                        & 0.25                        &0.27\\
$\theta_e$          & $-30\fdg2$             & $-36\fdg5$             & $-35\fdg9$             &$-37\fdg3$\\
$\gamma$            & 0.14                   & 0.087                       & 0.057                       &0.045\\
$\theta_{\gamma}$   & $ 38\fdg5$             & $50\fdg1 $             & $59\fdg3 $             &$66\fdg0$\\
$b_{G2}$            & --                     & $0\farcs63$                 & $0\farcs56$                 &$0\farcs65$\\
$\Delta\alpha_{G2}$ & --                     & $\equiv -1\farcs87$         & $\equiv -1\farcs87$         & $\equiv -1\farcs87$\\
$\Delta\delta_{G2}$ & --                     & $\equiv\phantom{+}4\farcs14$& $\equiv\phantom{+}4\farcs14$& $\equiv\phantom{+}4\farcs14$\\
$b_{\#1}$           & --                     &  --                         &  $0\farcs48$                &$0\farcs75$  \\
$\Delta\alpha_{\#1}$& --                     &  --                         & $\equiv\phantom{+}4\farcs50$&$\equiv\phantom{+}4\farcs50$\\
$\Delta\delta_{\#1}$& --                     &  --                         & $\equiv -6\farcs27$         &$\equiv -6\farcs27$\\
$b_{\#8}$           & --                     &  --                         &  --                         & $0\farcs036$  \\
$\Delta\alpha_{\#8}$& --                     &  --                         &  --                         &$\equiv\phantom{+}4\farcs25$\\
$\Delta\delta_{\#8}$& --                     &  --                         &  --                         & $\equiv\phantom{+}0\farcs61$\\
\tableline
\enddata
\footnotesize
\tablecomments{Position angles are measured East of
  North and positions are relative to image A (negative RA values are Eastward of Image A).  
  In rows 2-4, the $\chi^2$ is broken down into
  image positions ($\chi^2_{image}$), flux ratios ($\chi^2_{flux}$), G1 position
  ($\chi^2_{G1}$), and priors on $e$ and $\gamma$ ($\chi^2_{prior}$).}
\end{deluxetable}

Examining Table~\ref{tab:others}, we see that the next most important galaxies
    are \#1 and \#8, where adding an object near \#1 will also
    capture the primary effects of galaxies \#2 and H2.  If we
    add an SIS at the position of \#1 we find
    modest improvement in the fit, with $\chi^2=3.4$
    for $DOF=4$. Adding \#8 as a third galaxy does not improve the fit because its position is roughly orthogonal to the preferred 
shear direction, and we are using a prior of $\gamma = 0.05\pm0.05$ on the shear
    amplitude.  When adding both \#1 and \#8, we find $\chi^2=2.9$ for $DOF=3$. The results for these 3- and 4-galaxy
    models are summarized in Table~\ref{tab:2galmodelsH1413}. Note
    that the discrepancy between the best-fit critical radii in our
    models and the estimated critical radii in Table~\ref{tab:others}
    is not surprising given the uncertainties in the Faber-Jackson
    relation, in the H-band magnitudes, and in our best-fit critical radii. Part of
    the improvement in these multiple-galaxy models is due to a better
    fit to the flux ratios, but much is due to the reduced external
    shear and our shear prior. These results suggest we are seeing some effects from these
    other nearby galaxies, but we have also reached the limits of what we can measure
     given the available data.

\begin{deluxetable}{c c c c c c}
\tabletypesize{\footnotesize}
 \tablewidth{0pt}
\tablecaption{Lens Environment. \label{tab:others}}
\tablehead{
ID   & H$_n$ (mag) & $r_n (\arcsec)$& $b_n (\arcsec)$ & $\gamma_n$ & $b_{G1}b_n/r_n^2$}
\startdata
G1(=H1) & $20.6\pm0.5$ & --  & 0.5 & --   & --    \\
G2(=14) & $20.0\pm0.2$ & 4.0 & 0.6 & 0.08 & 0.019 \\ 
1       & $20.7\pm0.3$ & 8.3 & 0.5 & 0.03 & 0.003 \\
2       & $22.3\pm0.9$ & 7.3 & 0.2 & 0.02 & 0.002 \\
3       & $22.0\pm0.5$ & 8.3 & 0.3 & 0.02 & 0.002 \\
8       & $21.8\pm0.5$ & 4.4 & 0.3 & 0.03 & 0.007 \\
11      & $22.3\pm0.8$ & 7.3 & 0.2 & 0.01 & 0.002 \\
H2      & $21.6\pm0.7$ &11.0 & 0.3 & 0.01 & 0.001 \\
\tableline
\enddata
\footnotesize\tablecomments{H-band magnitudes (H$_n$), angular separations from G1,
  estimated critical radii, contributions to shear, and higher order, non-linear perturbations
  for nearby objects detected in H-band image of H1413+117 \citep{kne98a}.}
\end{deluxetable}

\section{Discussion}
\label{sec:summ}
A persistent embarrassment in lens models has been the inability
    to fit image flux ratios.  This is justified, in many cases
    correctly, by the complexities of the physical effects that
    modify flux ratios: absorption by the interstellar medium in
    the lens \citep[e.g.][]{fal99}, intrinsic time variability
    modulated by the time delays, gravitational microlensing
    by stars \citep[e.g.][]{wam06}, and more massive satellites
    or neighboring galaxies.  We know from the structure of
    the Einstein ring images of quasar host galaxies or lenses with other
    additional lensed features that the solution does not lie in
    giving the primary lens an odd angular structure \citep{yoo06,yoo06b,eva03,koc04,con05}.

    The advantage of mid-IR flux
    ratios is that they largely eliminate all of these complications
    other than the perturbing effects of satellites.  Given typical
    differential extinctions seen in lenses \citep{fal99}, the
    extinction corrections in the
    mid-IR are negligible. Quasars should show little mid-IR
    variability \citep[e.g.][]{poi07}, and the emission regions are too large to be
    affected by microlensing \citep[e.g.][]{ago00}.  It would, however, be wise to
    verify these latter two points empirically by coarsely monitoring
    a few lensed systems in the mid-IR.

    The remaining physical effect, perturbations from satellites (a.k.a.\
    CDM substructure), is thus well isolated.  To date, the failure of a
    central lens galaxy model to fit image positions has always been
    traced to a luminous perturbing object (e.g.\ MG0414+0534,
    \citealt{ros00}; MG2016+112, \citealt{mey06,mor09}; HE0435--1223, \citealt{koc06}).  That
    this is true for ``astrometric anomalies'' is not surprising
    because astrometric perturbations are dominated by the high-mass
    end of the satellite mass function which will tend to have formed
    stars.  It is also well-established that the flux-ratio anomalies can
    be explained by adding low-mass substructures \citep[see,
    e.g.][]{dal02,yoo06,con05,koo03,eva03,dob06,met02}, and the masses of
    the satellites can be so low that there is no surprise in
    a failure to observe the responsible object(s).  We should also
    note that the introduction of a visible satellite to
explain an astrometric anomaly does not necessarily explain any
accompanying flux ratio anomaly.  For example, while \citet{ros00} showed that a luminous satellite \citep{sch93} can explain the
image positions in MG0414+0534, the model still cannot explain
the flux ratio of the A$_1$ and A$_2$ images~-- a problem recently
confirmed by the mid-IR observations of \citet{min09}.

Nonetheless,
    high-mass substructures should also produce flux ratio anomalies,
    and we should be able to demonstrate that these can be correctly
    detected and attributed.   While our original goal was to use the mid-IR flux ratios to search 
for evidence of low-mass, non-luminous substructure in the lens 
H1413+117, we instead found that the flux ratios could be explained
by a nearby galaxy.  This galaxy, which we refer to as G2, is object
\#14 in the Kneib et al.\ (1998ab) catalog, and it lies $1\farcs87$ East and
$4\farcs15$ North of image A.  It has comparable luminosity to the main
lens, and it is close enough to the lens for the higher order terms
in its perturbations to the lens potential beyond an external shear to
significantly modify the image magnifications.  Adding other nearby
galaxies does not significantly improve the fits, in large part because
we have a reduced $\chi^2$ near unity after adding G2.  More 
importantly, no perturbing galaxies are required to explain the
astrometry~--~the detection of the neighboring galaxy is purely
due to the flux ratios. The only similar
    claim is for the four-image lens B2045+265,
    where the anomaly can be explained by turning a relatively round
    visible satellite into a 10:1 flattened mass distribution running
    across the most anomalous image \citep{mck07}. While the model works to explain
    the anomaly, the mass distribution assigned to the
    satellite is not very physical.

Independent of any such debates, mid-IR imaging provides a
    simple means of greatly expanding the numbers of lenses where
    these issues can be studied quantitatively. While there are
    10 known radio lenses with four or more images, there are
    17 such optical lenses. Moreover, radio flux ratios must also be corrected
      for intrinsic variability modulated by the time delays
    \citep[e.g.][]{fas02}, while there is at least
      an expectation that these effects will be minimal in the mid-IR.
      If true, single epoch measurements suffice in the mid-IR, while
      monitoring campaigns are required for radio wavelengths. Mid-IR flux ratios
    will also help in models for microlensing, since a better mass model can be constructed, and the
microlensing flux anomalies will be better measured.  Furthermore, a survey of
microlensing comparing the optical to the mid-IR (assuming the optical can be
corrected for extinction) could place a constraint on the ratio of
    dark matter to stellar matter at the position of the lensed images
    \citep{sch02}.

Unfortunately, little use has been made of this approach. 
      There are published measurements for Q2237+0305 \citep{ago00,min09}, 
B1422+231 and PG1115+080 \citep{chi05}, MG0414+0534 \citep{min09}, and now
    our measurements for H1413+117.  All
      of these measurements are of modest precision (5--10\%), so
      there is considerable scope both for expansion of the sample
      and improvements in the precision.  Spitzer/IRAC  observations
      (e.g.\ HE1104--1805, \citealt{poi07}; SDSS1004+4112,
       Ross et al.\ in prep), have much higher precision but can only
      resolve the larger separation lenses due to the poor 
resolution of Spitzer (FWHM of $1\farcs6$ and $1\farcs9$ at 3.6 and $8.0\mu$m,
      respectively). Moreover, the
highest resolution bands will be dominated by more compact accretion disk emission
in the rest-frame of most quasars (3.6$\mu$m would be $1\mu$m in
      the rest-frame of H1413+117).  Thus, further progress will
      depend on ground-based observations until the advent of the
    James Webb Space Telescope (JWST), which will provide
    exquisite mid-IR precision for every known gravitational lens.
    These observations are challenging for the 
current generation of 8m telescopes, but will become relatively easy
with the next generation of 30m telescopes.

\acknowledgments

This work is based on observations obtained at the Gemini Observatory
(with Program ID GN-2005B-Q-43), which is operated by the
Association of Universities for Research in Astronomy, Inc., under a cooperative agreement
with the NSF on behalf of the Gemini partnership: the National Science Foundation (United
States), the Science and Technology Facilities Council (United Kingdom), the
National Research Council (Canada), CONICYT (Chile), the Australian Research Council
(Australia), Ministério da Ciência e Tecnologia (Brazil) 
and Ministerio de Ciencia, Tecnología e Innovación Productiva
(Argentina). This work is based in part on observations made with the NASA/ESA Hubble 
Space Telescope. Support for program GO-7495 was provided by NASA 
through a grant from the Space Telescope Science Institute, which is
operated by AURA, Inc., under NASA contract NAS5-2655.  EA is
partially supported by National Science Foundation CAREER Grant No.
0645416. CSK is supported by NSF grant AST-0708082.

{\it Facilities:} \facility{Gemini (Michelle)}, \facility{HST (NICMOS)}.

\end{document}